\documentclass[a4paper,11pt]{article}
\usepackage{pos}
\usepackage{amsmath}
\usepackage{setspace}

\def\approxprop{%
  \def\p{%
    \setbox0=\vbox{\hbox{$\propto$}}%
    \ht0=0.6ex \box0 }%
  \def\s{%
    \vbox{\hbox{$\sim$}}%
  }%
  \mathrel{\raisebox{0.7ex}{%
      \mbox{$\underset{\s}{\p}$}%
    }}%
}

\title{Observational constraints on the maximum energies of accelerated particles in supernova remnants}

\author*[a, b]{Hiromasa Suzuki}
\author[b, c]{Aya Bamba}
\author[d, e]{Ryo Yamazaki}
\author[f]{Yutaka Ohira}

\affiliation[a]{Department of Physics, Faculty of Science and Engineering, Konan University, 8-9-1 Okamoto, Higashinada, Kobe, Hyogo 658-8501, Japan}
\affiliation[b]{Department of Physics, The University of Tokyo, 7-3-1 Hongo, Bunkyo-ku, Tokyo 113-0033, Japan}
\affiliation[c]{Research Center for the Early Universe, The University of Tokyo, 7-3-1 Hongo, Bunkyo-ku, Tokyo 113-0033, Japan}
\affiliation[d]{Department of Physics and Mathematics, Aoyama Gakuin University, 5-10-1 Fuchinobe, Chuo-ku, Sagamihara, Kanagawa 252-5258, Japan}
\affiliation[e]{Institute of Laser Engineering, Osaka University, 2-6 Yamadaoka, Suita, Osaka 565-0871, Japan}
\affiliation[f]{Department of Earth and Planetary Science, The University of Tokyo, 7-3-1 Hongo, Bunkyo-ku, Tokyo 113-0033, Japan}

\emailAdd{hiromasa050701@gmail.com}

\abstract{Supernova remnants (SNRs) are thought to be the most plausible sources of Galactic cosmic rays. One of the principal questions is whether they are accelerating particles up to the maximum energy of Galactic cosmic rays ($\sim$PeV). In this paper, we summarize our recent studies on gamma-ray-emitting SNRs.
We first evaluated the reliability of SNR age estimates to quantitatively discuss time dependence of their acceleration parameters. Then we systematically modeled their gamma-ray spectra to constrain the acceleration parameters. The current maximum energy estimates were found to be well below PeV for most sources. The basic time dependence of the maximum energy assuming the Sedov evolution ($\approxprop t^{-0.8\pm0.2}$) cannot be explained with the simplest acceleration condition (Bohm limit) and requires shock-ISM (interstellar medium) interaction. The inferred maximum energies during lifetime averaged over the sample can be expressed as $\lesssim 20$ TeV ($t_{{\rm M}}/\text{1~kyr})^{-0.8}$ with $t_{\rm M}$ being the age at the maximum, which reaches $\sim$PeV only if $t_{\rm M} \lesssim 10$~yr. The maximum energies during lifetime are suggested to have a variety of 1--2 orders of magnitude from object to object on the other hand. This variety will reflect the dependence on environments.
}

\FullConference{%
  7th Heidelberg International Symposium on High-Energy Gamma-Ray Astronomy (Gamma2022)\\
  4-8 July 2022\\
  Barcelona, Spain\\}

\begin{document}
\maketitle

\section{Introduction}

Galactic cosmic rays are high-energy particles with an energy spectrum approximated by a power-law and a maximum energy of $\approx 10^{15.5}$ eV ($\approx 3$~PeV).
Their acceleration sites are still unclear despite of more than 100~yr investigations.
Supernova remnants (SNRs) are thought to be the most promising acceleration sites that provide these high-energy particles.
According to analytical models for diffusive shock acceleration, these sources are believed to actually supply particles with energies of $\lesssim$~PeV (e.g., \cite{bell78, lagage83}).

Gamma-ray observations have revealed that charged particles are accelerated to energies above TeV (10$^{12}$~eV) in young SNRs.
Most of these SNRs feature, however, spectral turnovers at energies below PeV.
Recently, Tibet Air-Shower array and LHAASO (Large High Altitude Air Shower Observatory) observations have identified a number of PeV accelerators (PeVatrons) \citep{amenomori21, cao21}, which includes only a few SNRs.
These results suggest that PeV particles accelerated in early evolutionary phases have already escaped (e.g., \cite{ptuskin03, ohira10, schure13, cardillo15, celli19, inoue21}), or that there are almost no PeVatron SNRs.
Other scenarios have also been proposed in which SNRs in specific conditions become PeVatrons: e.g., very young SNRs in dense environments \cite{cristofari20}.


In this paper, we summarize our recent observational works on investigation of the maximum attainable energy of SNRs.
In Section~\ref{sec-age}, we quantitate the reliability of age estimates \citep{suzuki21a}.
Section~\ref{sec-gamma} summarizes our systematic analysis on the gamma-ray spectra of 38 sources \citep{suzuki22a}.
Using the age and gamma-ray properties, we extract the basic time dependence and variety of the maximum attainable energies in Section~\ref{sec-max}.
Our basic conclusion is that SNRs can be PeVatrons only if they accelerate particles very efficiently when they are very young such as less than 10~yr (Section~\ref{sec-sum}).

\section{Reliability of supernova remnant age estimates}\label{sec-age}
In order to discuss time dependence of particle-acceleration properties quantitatively, we first evaluate the uncertainties in SNR age estimation using systems with reliable age measurements $t_{\rm r}$, either the historical age, light-echo age, {$t_{\rm kin, ej}$} (kinematic age of free-moving ejecta knots), or $t_{\rm kin, NS}$ (kinematic age of associated neutron star) \cite{suzuki21a}.
First, we measure $t_{\rm kin, NS}$ for available systems by estimating the geometric centers of the SNR shells. The velocities of neutron stars are taken from multi-epoch radio observations.
Then we ``calibrate'' the general age estimates, which are applicable for most cases\footnote{Dynamical age $t_{\rm dyn}$ and plasma age $t_{\rm p}$}, by comparing them with the reliable estimates $t_{\rm r}$.

The sample for this section consists of the SNRs with known historical or light-echo ages, {\bf $t_{\rm kin, ej}$}, or measurable $t_{\rm kin, NS}$.
The best age $t_{\rm b}$ is defined for each source, which is the most reliable estimate.
Details of the sample are explained in \cite{suzuki21a}.

Ideally, where a symmetric supernova explosion occurs in a uniform ambient density, the SNR will have a circular shell and so the geometric center can be easily determined.
In reality, however, the observed SNR shapes are not simply circular. Thus, we model the projected morphology of the SNR shell with an ellipse as a simple enough yet better alternative to a circle and estimate the geometric center.
Then we can calculate $t_{\rm kin, NS}$ using the neutron star position and velocity.

\begin{figure*}[htb]
\centering
\includegraphics[width=14cm, angle=0]{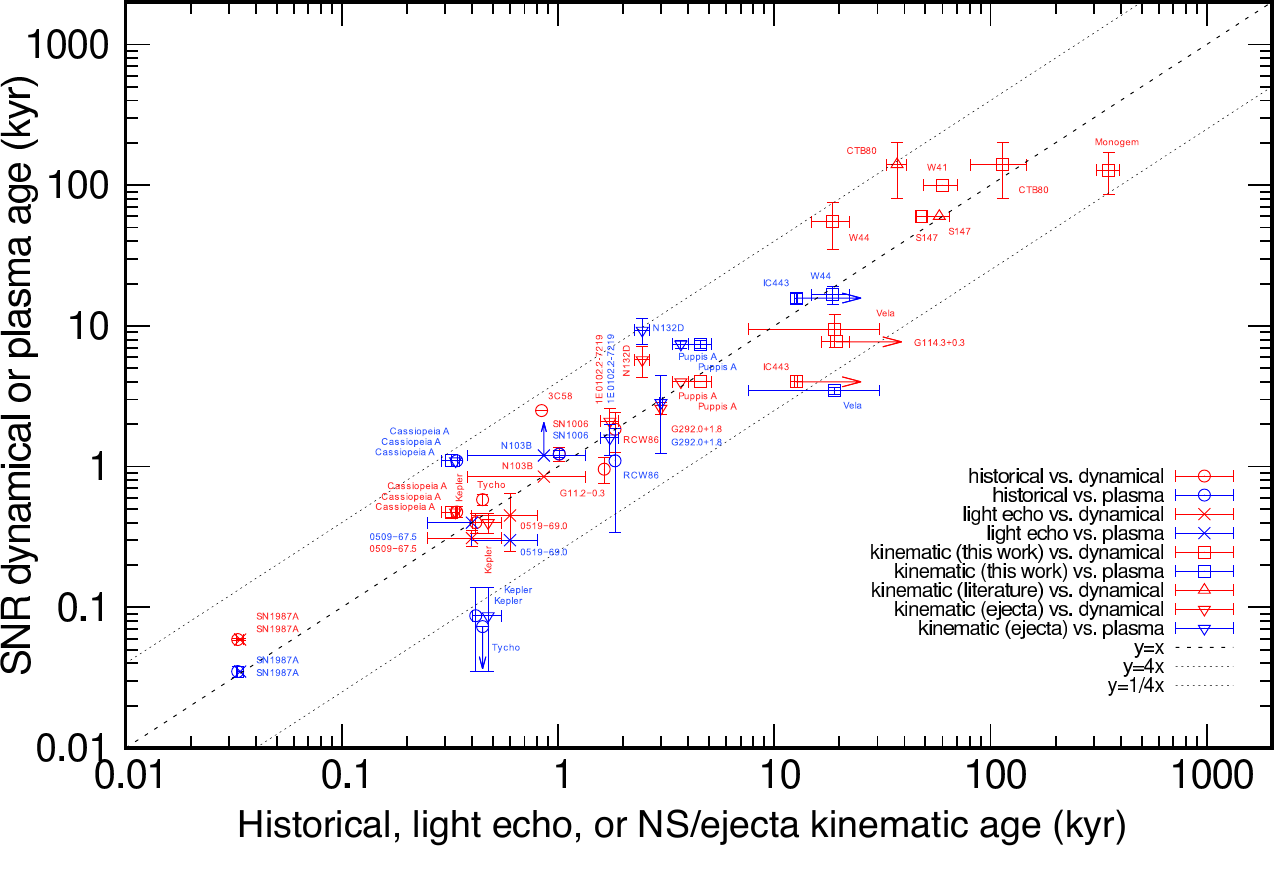}
\caption{Comparison of a variety of age estimates for SNRs.
The $x$-axis represents reliable estimates $t_{\rm r}$ (circles for historical ages, crosses for light-echo ages, inverted triangles for $t_{\rm kin, ej}$, squares for $t_{\rm kin, NS}$, triangles for $t_{\rm kin, NS, prev}$).
The $y$-axis represents the dynamical (red) or plasma (blue) ages.
For reference, the linear functions $y=x, 4x$, and $1/4\, x$ are also plotted.
\label{fig-ages1}}
\end{figure*}

Figure~\ref{fig-ages1} plots $t_{\rm dyn}$ and $t_{\rm p}$ as a function of $t_{\rm r}$ (historical ages, light echo ages, and/or $t_{\rm kin, NS}$).
Ideally, all the points would be on the straight line on the $x$-$y$ plane ($y = x$).
The $t_{\rm dyn}$ and $t_{\rm p}$ values of most of the SNRs in our sample are found to be in good agreement with their $t_{\rm r}$ within a factor of four.
So, we tentatively conclude that a systematic error of a factor of four is associated with the general age estimates.

\section{Gamma-ray spectral properties}\label{sec-gamma}
\begin{figure*}[h!]
\centering
\includegraphics[width=16cm, angle=0]{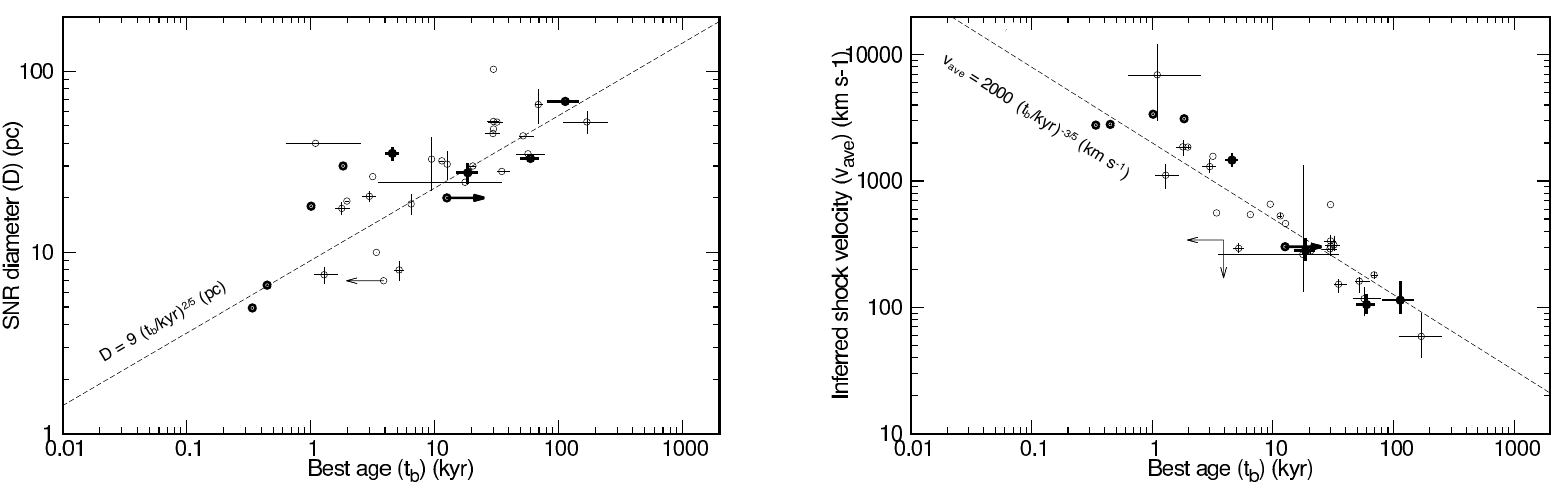}
\caption{
Left and right panels show the SNR diameter $D$ and inferred average shock velocity $v_{\rm ave}$ as a function of $t_{\rm b}$, respectively.
A condition $v_{\rm ave} = D/5t_{\rm b}$ is assumed. Note that the objects shown here are only those with gamma-ray emissions.
A representative Sedov model, satisfying the condition $D = C_D t^{2/5}$ with the parameter $C_D$ selected by eye to roughly match the data, is also plotted with black-dashed lines.
\label{fig-diameter-velocity}}
\end{figure*}

Here we describe the essence of our analysis and results performed in \cite{suzuki22a}.
Details of the sample selection can be found in \cite{suzuki22a}.
Figure~\ref{fig-diameter-velocity} shows the diameter $D$ and inferred shock velocity $v_{\rm ave}$ as a function of $t_{\rm b}$.\footnote{We calculate this velocity assuming the Sedov model as $v_{\rm ave} = D/5t_{\rm b}$.}
A representative Sedov model, for which a condition $D = C_D t^{2/5}$ is assumed and the parameter $C_D$ is selected by eye to roughly match the data, is also plotted in Figure~\ref{fig-diameter-velocity}.
Both the $D$--$t_{\rm b}$ and $v_{\rm ave}$--$t_{\rm b}$ plots are close to the representative Sedov model, indicating that the estimated age is plausible.

In our work, hadronic emission is assumed to dominate the gamma rays, which seems to be the case at least for several SNRs based on their spectral shapes and energetics (e.g., \cite{ackermann13}).
The exponential cutoff energy ($E_{\rm cut}$) or break energy ($E_{\rm br}$) of a gamma-ray spectrum reflect the maximum energies of fresh accelerated particles \citep{ohira11a, celli19, brose20}.
If particles no longer accelerated do not significantly contribute to the emission, an exponential-like cutoff feature is expected.
On the other hand, if the contribution of escaping particles cannot be ignored, the gamma-ray spectrum will be similar to a broken power-law.
Since generally we cannot distinguish between these two situations without detailed information on acceleration sites, the gamma-ray spectra are fitted with both an exponential cutoff power-law model and a broken power-law model.
By fitting the Fermi, H.E.S.S., MAGIC, and VERITAS spectra, we obtain parameters such as spectral index, cutoff, break, and hardness ratio.


\section{Time dependence of maximum acceleration energy and variety}\label{sec-max}
\begin{figure*}[htb!]
\centering
\includegraphics[width=16cm, angle=0]{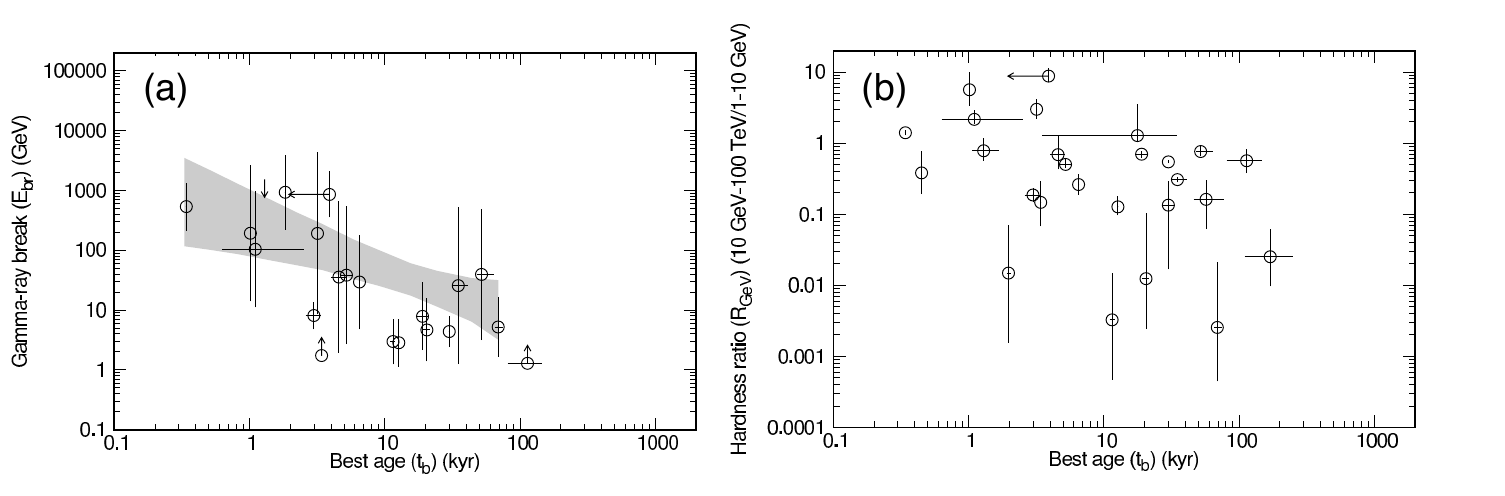}
\caption{Plots of gamma-ray break energy (a) and hardness ratio (b) over age.
Grey regions represent the best-fit power-law functions and their $1\sigma$ confidence ranges.
\label{fig-cbhl}}
\end{figure*}

Figures~\ref{fig-cbhl} (a) and (b) show $E_{\rm br}$ and hardness ratio\footnote{Flux ratio, 10~GeV--100~TeV over 1--10~GeV.} over age, respectively.
Both of them show general decreasing trends with age with large scatters, which are consistent with previous works \cite{zeng19, suzuki20b}.
The $E_{\rm cut}$--$t_{\rm b}$ and $E_{\rm br}$--$t_{\rm b}$ plots are modeled with a power-law model, and the best-fit functions are obtained as $E_{\rm cut} = 1.3_{-0.67}^{+1.1}~{\rm TeV}~ (t_{\rm b} / 1~{\rm kyr})^{-0.81 \pm 0.24}$ and $E_{\rm br} = 270_{-130}^{+240}~{\rm GeV}~ (t_{\rm b} / 1~{\rm kyr})^{-0.77 \pm 0.23}$.
Thus, based on our simple extrapolation with an assumption of the Sedov evolution, only SNRs younger than $\sim 10$~yr have a chance to become PeVatrons. This scenario matches some theoretical works \cite{cardillo15, cristofari20, inoue21}.
The time dependence of $\approxprop t^{-0.8}$ is different from that expected in the simplest regime, Bohm limit ($t^{-0.2}$) \cite{vink20}, and requires wave damping via shock-ISM (interstellar medium) collisions (e.g., \cite{ptuskin03, ptuskin05, yasuda19, brose20}\footnote{This discrepancy will be partly due to the assumption of the Sedov evolution for all the sources.}.
We also find a large variety of the maximum energy estimates even at the same ages: we quantify this variety to be 1--2 orders of magnitude.

\begin{figure}[htb!]
\centering
\includegraphics[width=14cm]{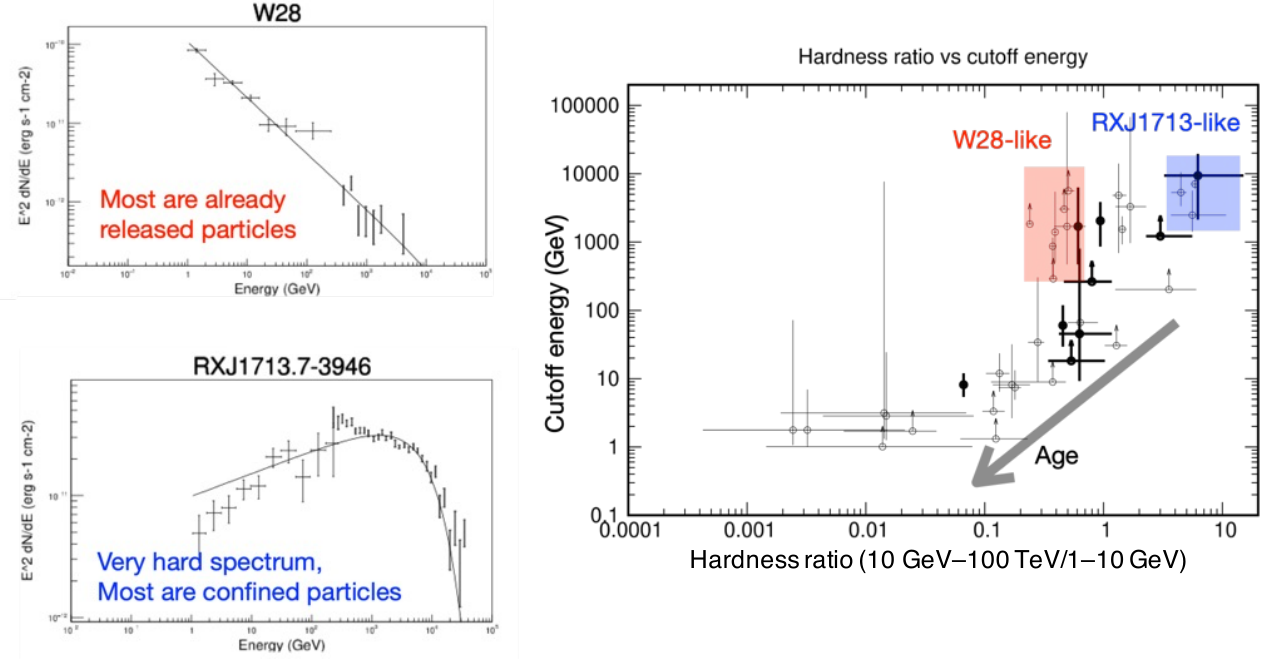}
\caption{
Scatter plot of gamma-ray cutoff energy over hardness ratio (right) and two example gamma-ray spectra (left) extracted from two outlier groups, W28-like and RXJ1713-like objects.
\label{fig-hardness-cutoff}}
\end{figure}

In the plot of the cutoff energy over hardness ratio in Figure~\ref{fig-hardness-cutoff}, most sources show a similar trend evolving from top right to bottom left, namely young systems show a hard emission.
There are some outliers, however, which can be classified as ``W28-like'' and ``RXJ1713-like'' objects. The maximum energy estimates are well constrained for these outliers because the contributions of fresh accelerated particles and escaping particles can be separated well (refer to \cite{suzuki22a} for details). Theoretical studies suggest that these peculiar properties are due to complex gas environments \cite{nava13, celli19}.

\section{Summary}\label{sec-sum}
In our recent studies, we investigated the basic time dependence and variety of acceleration properties of SNRs.
We first calibrated general age estimates using reliable estimates available only for specific systems.
As a result, the general estimates of the SNRs were found to agree with $t_{\rm r}$ within a factor of four.
A systematic analysis of 38 gamma-ray emitting SNRs using their thermal X-ray and gamma-ray properties was performed.
A spectral modeling on their gamma-ray spectra has allowed us to constrain the particle-acceleration parameters.
Two candidates of the maximum energy of freshly accelerated particles, the gamma-ray cutoff and break energies, were found to be well below PeV for our sample.

The general time dependences of the maximum energy estimates for our sample, $\approxprop t^{-0.8}$, cannot be explained with the simplest acceleration condition of the Bohm limit.
The estimated average maximum energies of accelerated particles during lifetime $\lesssim 20$~TeV $(t_{\rm M}/1~{\rm kyr})^{-0.8}$ are well below PeV if the age at the maximum $t_{\rm M}$ is $\sim 100$--1000~yr.
However, if the maximum energy during lifetime is realized at younger ages such as $t_{\rm M} < 10$~yr, it can become higher and reach PeV.
On the other hand, the maximum energies during lifetime are suggested to have a large variety of 1--2 orders of magnitude from object to object.
This will reflect the dependence of the acceleration processes on environments.

\small
\setstretch{0.8}
\bibliography{/Users/suzuki/Documents/paper_submission/references/references.bib}
\bibliographystyle{unsrt}

\end{document}